\documentstyle[12pt,epsf]{article}
 \newcommand{\R}{{\rm I \hspace{-0.52ex} R}}
  \newcommand{\N}{{\rm I \hspace{-0.52ex} N}}
\begin{document}
\setlength\textheight{8.75in}
\newcommand{\be}{\begin{equation}}
\newcommand{\ee}{\end{equation}}
\title{On linear operators with an invariant subspace of functions}
\author{{\large Yves Brihaye\footnote{ yves.brihaye@umh.ac.be}} \\
{\small Facult\'e des Sciences, Universit\'e de Mons-Hainaut, }\\
{\small B-7000 Mons, Belgium }\\
}

\date{\today}
\maketitle
\thispagestyle{empty}

\begin{abstract}
Let us denote ${\cal V}$, the
finite dimensional vector spaces of functions of the form 
$\psi(x) = p_n(x) + f(x) p_m(x)$
where $p_n(x)$ and $p_m(x)$ are arbitrary  polynomials of degree
at most $n$ and $m$ in the variable $x$ while $f(x)$ represents a
fixed  function of $x$.
 Conditions on $m,n$ and $f(x)$ are found  such that  families of
linear differential operators exist which preserve ${\cal V}$. A special 
emphasis is accorded to the cases where the set of differential operators
represents the envelopping algebra of some abstract algebra.
\end{abstract}
\medskip
\medskip
\newpage
\section{Introduction}
Quasi Exactly Solvable (QES) operators are characterized by 
linear differential operators
which preserve a finite-dimensional vector space ${\cal V}$  
of smooth functions \cite{tur1}.
In the case of operators
of one real variable the underlying vector space is often of the form 
${\cal V} = {\cal P}_n$ where
${\cal P}_n$  represents the vector space of polynomials of degree 
at most $n$ in the variable $x$.
In \cite{tur2} it is shown that the linear operators preserving 
${\cal P}_n$ are generated by three
basic operators $j_-, j_0 , j_+$
(see Eq.(\ref{j}) below) which realize the algebra $sl(2,\R)$.
More general QES operators
can then be constructed by considering the elements of the enveloping
algebra of these generators, performing a change of variable 
and/or conjugating
the $j$'s with an invertible function, say
$g(x)$.  The effective invariant space is then the set of functions
of the form $g(x){\cal P}_n$.

In this paper we consider a more general situation.
 Let $m,n$ be two positive integers
  and let $f(x)$ be a sufficiently derivable function
 in a domain of the real line.
 Let ${\cal V}={\cal P}_n + f(x){\cal P}_m$
 be the vector space of functions of the form
 $p(x) + f(x) q(x)$  where $p(x)\in {\cal P}_n$,
 $q(x) \in {\cal P}_m$.

 We want to address the following questions~:
 What are the differential operators which preserve ${\cal V}$  ?
 and for which choice of $m,n,f(x)$ do these operators posses a relation
 with the enveloping algebra of some Lie (or "deformed`` Lie) algebra ?.
 This question generalizes the cases of monomials
 adressed in \cite{tur3} and more recently in \cite{dv}
At the moment  the question is, to our knowledge, 
not solved in its generality
but we present a few non trivial solutions in the next section.

\section{Examples}
\subsection{Case $f(x) = 0$}
This is off course the well known case of \cite{tur1,tur2}. The
relevant operators
read
\be
\label{j}
     j_+(n) = x( x \frac{d}{dx} - n) \ \ , \ \
     j_0(n) = ( x \frac{d}{dx} - \frac{n}{2}) \ \ , \ \
     j_- = \frac{d}{dx}
\ee
and represent the three generators of $sl(2,\R)$. Most of known
one-dimensional QES equations are build with these operators.
For later convenience, we further define a family of equivalent
realizations of $sl(2,\R)$  by means of the conjugated operators
$k_{\epsilon}(a)\equiv x^a j_{\epsilon} x^{-a}$
for $\epsilon = +,0,-$ and $a$ is a real number.

\subsection{Case $f(x) = x^a$}
The general cases of vector spaces constructed over monomials
was first adressed
in \cite{tur3} and the particular subcase $f(x) = x^a$ was
reconsidered recently \cite{dv}.
The corresponding vector space  was
denoted $V^{(1)}$ in \cite{dv};
here we will reconsider this case and extend the discussion
of the operators
which leave it invariant. For later convenience,
it is usefull to introduce more precise notations, setting
${\cal P}_n \equiv {\cal P}(n,x) $ and
\begin{eqnarray}
\label{v1}
V^{(1)} \equiv  V^{(1)}(N,s,a,x) &=& {\cal P}(n,x)
+ x^a {\cal P}(m,x) \nonumber \\
&=& {\rm span} \{1,x,x^2,\dots ,x^n; x^a, x^{a+1}, \dots ,x^{a+m} \}
\nonumber \\
&=& V_1^{(1)} \oplus V_2^{(1)}
\end{eqnarray}
in passing, note that the notations of
\cite{dv} are  $n = s$ and $m = N-s-2$.

The vector space above is clearly constructed as the
direct sum of two subspaces.
As pointed out in \cite{tur3,dv}
three independent, second order differential operators
can be constructed which preserve
the vector space $V^{(1)}$. Writing these operators in the form
\begin{eqnarray}
      &J_+ &= x (x \frac{d}{dx} - n)
      (x \frac{d}{dx} - (m+a))  \nonumber \\
\label{jj}
      &J_0 &=  (x \frac{d}{dx} - \frac{m+n+1}{2})    \\
      &J_- &=  (x \frac{d}{dx} + 1 - a) \frac{d}{dx} \nonumber
\end{eqnarray}
makes it obvious that they preserve  $V^{(1)}$.

These operators close under the commutator
into a polynomial deformation of the $sl(2,\R)$ algebra:
\begin{eqnarray}
\label{nlalgebra}
 &[ J_0 , J_{\pm}] &= \pm J_{\pm} \nonumber \\
 &[ J_{+} , J_{-}] &= \alpha J_0 ^3 + \beta J_0 ^2 + \gamma J_0 + \delta
 \end{eqnarray}
where $\alpha, \beta, \gamma, \delta$ are constants  given in \cite{dv}.

Clearly the operators (\ref{jj}) leave
separately invariant  two vector spaces
$V_1^{(1)}$ and  $V_1^{(1)}$ entering in (\ref{v1}).
In the language of representations  they act reducibly on $V^{(1)}$.
However,  operators can be constructed which
preserve $V^{(1)}$ while
mixing the two subspaces. The form of these supplementary operators
is different according
to the fact that the number $a$ is an integer or not; 
we now adress these two cases
separately.
\subsubsection{ $a\in \R$  }
In order to construct the operators which mix
$V_1^{(1)}$ and $V_2^{(1)}$, we first define
\begin{eqnarray}
&K &= (D-n)(D-n+1) \dots D  \ \ , \ \ D \equiv x\frac{d}{dx} \nonumber \\
&K' &= (D-m-a)(D-m-a+1) \dots (D-a)
\end{eqnarray}
which  belong to the kernals of the subvector spaces
${\cal P}_n$ and  $x^a {\cal P}_m$ of  $V^{(1)}$ respectively.
Notice that the products $j_\epsilon \tilde K$ and
$k_{\epsilon}(a) K$ (with $\epsilon = 0, \pm$)
also preserve the vector space.  For generic values
of $m,n$ these operators contain more than second derivatives
and, as so, they were not considered in \cite{dv}.

In order to construct the operators which
mix the two vector subspaces entering in ${\cal V}$,
we first have to construct the operators which transform a generic
element of ${\cal P}_m$ into an element of ${\cal P}_n$
and vice-versa.
 In \cite{bk} it is shown that these operators are of the form
\begin{eqnarray}
&q_{\alpha} &= x^{\alpha} \ \ \ , \ \ \alpha = 0,1,\dots, \Delta \ \\
&\overline q_{\alpha}&= \prod_{j=0}^{\alpha-1}(D-(p+1-\Delta)-j)
(\frac{d}{dx})^{\Delta-\alpha}
\end{eqnarray}
 where $\Delta \equiv \vert m-n \vert$ , $p \equiv $max$\{m,n\}$

The operators preserving ${\cal V}$ while exchanging the two
subspaces can finally be constructed by means of
\be
Q_{\alpha} = q_{\alpha} x^{-a} K \ \ , \ \
\overline Q_{\alpha} =  x^a \overline q_{\alpha}  K' \ \  , \ \
\alpha = 0,1,\dots, \Delta.
\ee
Here we assumed $n \leq m$, the case $n \geq m$ is
obtained by exchanging $q_{\alpha}$ with $ \overline q_{\alpha}$
in the formula above.

 It can be checked easily that
 $Q_{\alpha}$, transform a vector of the form $p_n + x^a q_m$ into
 a vector of the form $\tilde q_n \in {\cal P}_n$
 while  $\overline Q_{\alpha}$
 transforms the same vector
 into a vector of the form  $x^a \tilde p_m \in x^a {\cal P}_m$.

The generators constructed above are in one to one
correspondance with the 2$\times$2 matrix generators
preserving the direct sum of vector spaces
${\cal P}_m \oplus {\cal P}_n$ classified in \cite{bk}
although  their form is quite different (the same
notation is nevertheless used).
The commutation relations
(defining a normal order) which the generators fullfill is also
drastically different as we shall
discuss now.
First of all it can be easily checked that all products of
operators $Q$ (and separately of $\overline Q$)
belong to the kernal of the full space $V^{(1)}$, so we can write
\be
     Q_{\alpha} Q_{\beta} = \overline Q_{\alpha} \overline Q_{\beta} = 0
\ee
which suggests that the operators $Q$'s and the $\overline Q$'s
play the role of fermionic generators,
in contrast to the $J$'s which  are bosonic
(note that the same distinction holds in the case \cite{bk}).

From now on, we assume $n=m$ in this section
(the evaluation of the commutators for generic values
of $m,n$ is straightforward but leads to even more involved
expressions)
and suppress the superflous index $\alpha$ on the
the fermionic operators. The commutation relations between
fermionic and bosonic generators leads to
\begin{eqnarray}
   &[Q,J_-] = (2a-n-1) j_- Q \  , \
   &[Q,J_+]=(2a+n+1) j_+ Q    \\
   &[\overline Q,J_-] = -(2a+n+1) k_-(a) \overline Q \  ,  \
   &[\overline Q,J_+]= -(2a-n-1) k_+(a) \overline Q  \nonumber
\end{eqnarray}
where the $j_{\pm}$ and $k_{\pm}(a)$ are defined in (\ref{j}).
These relations define a normal order but we notice that the right
hand side are not linear expressions of the generators choosen
as basic elements.
We also have
\be
        [Q, D ]  = (D+a) Q  \ \   , \ \
        [\overline Q, D]  = (D-a) \overline Q
\ee
This is to be contrasted with the problem studied in \cite{bk}
where, for the case $\Delta = 0$,
the $Q$ (and the $\overline Q$)
commute with the three bosonic generators,
forming finally an sl(2)$\times$ sl(2) algebra.
Here we see that the bosonic operators $J$ and
fermionic operators $Q, \overline Q$
do not close linearly under the commutator.
The commutators involve in fact extra factors which can
be expressed in terms of the
operators $j$ or $k(a)$ acting on the appropriate subspace
${\cal P}_m$. This defines a normal order among
the basic generators but makes the underlying algebraic
stucture (if any) non linear.  For completeness, we also
mention that the anti-commutator
$\{Q, \overline Q \}$ is a polynomial in $J_0$.
 \subsubsection{$a \in \N$}
 Let us consider the case $a \equiv k \in \N_0$, with
 $n \leq k $ and assume for definiteness
 $ m-k \geq n$.
 Operators that preserve $V^{(1)}$ while exchanging
 some monomials of the subspace $V_1^{(1)}$  with some of
 $V_2^{(1)}$  (and vice versa) can be expressed as follows~:
 \be
    W_+ = x^k \prod_{j=0}^{k-1} ( D - k - m + j)
 \ee
 \be
    W_- = \frac {1}{x^k} \prod_{j=0}^{n} ( D - j)
          \prod_{i=1}^{k-n-1} (D - k - n - i)
 \ee
 These operators are both of order $k$, $W_+$ is of degree $k$
 while $W_-$ is of degree $-k$.
 When acting  on the monomial of Eq.(\ref{v1}), $W_+$
  transforms the $n+1$ monomials of $V_1^{(1)}$
 into the first $n+1$ monomials of $V_2^{(1)}$ and annihilates the
 $k$ monomials of highest degrees in $V_2^{(1)}$.  To the contrary
 $W_-$ annihilates the $n+1$ monomials of $V_1^{(1)}$ and shifts
 the $n+1$ monomials of lowest degrees of   $V_2^{(1)}$  into
 $V_1^{(1)}$.  Operators of the same type performing higher jumps
 can be constructed in a straighforward way;
 they are characterized by a higher order and higher
 degrees but we will not present them
 here.  
 
 The two particular cases $k=n+1$ and $k=2$ can be further commented.
 In the case $k=n+1$ the space $V^{(1)}$ is just ${\cal P}_{m+n+1}$
 and the operators $W_+$, $W_-$ can be rewritten as
 \be
        W_+ = (j_+(m+n+1))^{n+1}   \ \ , \ \ W_- = (j_-)^{n+1}
 \ee
where $j_{\pm}$ are defined in (\ref{j}).
 Setting $k=2$ (and $n=0$ otherwise we fall on the case just mentionned),
 we see that the operators $W_{\pm}$ become
 second order  and coincide with the operators noted
 $T_2^{(+2)}$ ,  $T_2^{(-2)}$ in Sect. 4
 of the recent preprint \cite{guk}. With our notation they read
 \be
       W_+ = x^2(D-(m+2))(D-(m+1)) \ \ , \ \ 
       W_- = x^{-2} D (D-3)
 \ee

 A natural question which come out is to study whether  the
  non-linear algebra (\ref{nlalgebra}) is extended
  in a nice way by the supplementary
  operators $W_{\pm}$ and their higher order
  counterparts. So far, we have not found any interesting
  extended structure.
  For example, for $k=2,n=0$ , we computed~:
  \begin{eqnarray}
  &[W_+ , J_+] &= -2 x^3 (D-(m+2))(D-(m+1))(D-m) \nonumber \\
  &[W_+ , J_-] &= -6 x D (D-(m+2)) (D-\frac{2}{3}(m+2))
   \end{eqnarray}
 which just show that the commutators close within the envelopping
 algebra of the $V^{(1)}$ preserving operators
 but would need more investigation to be confirmed as an abstract
 algebraic stucture.

We end up this section by  mentionning that
the two other vector spaces constructed in \cite{dv} and
denoted $V^{(a-1)}$ and $V^{(a)}$
can in fact be related to $V^{(1)}$
by means of the following relations~:
\be
V^{(a-1)}(x)   = V^{(1)} (N, s=0 , \frac{1}{a-1} , x^{a-1})
\ee
\be
V^{(a)}(x)   = V^{(1)}   (N, s, \frac{1}{a} , x^{a} )
\ee
Off course the operators  preserving them
can be  obtained
from the operators above (\ref{jj}) after
a suitable change of variable and the results above
can easily be extended to  these vector spaces.

\subsection{Case $f(x) = \sqrt{p_2(x)}$, $m=n-1$}
Here, $p_2(x)$ denotes a polynomial of degree 2 in $x$, we
take it in the canonical form  $p_2(x) = (1-x)(1-\lambda x)$.
In the case $m = n-1$, three basic operators can be contructed which
preserve ${\cal V}$; they are of the form
the form
\begin{eqnarray}
\label{sop}
       S_1 &=& n x + p_2 \frac{d}{dx}    \nonumber \\
       S_2 &=& \sqrt{p_2}(n x - x \frac{d}{dx})     \\
       S_3 &=& \sqrt{p_2}(\frac{d}{dx})  \nonumber
\end{eqnarray}
and obey the commutation relations of so(3).  The family
of operators preserving ${\cal V}$ is in this case the enveloping
algebra of the Lie algebra of SO(3) in the realization above.
Two particular cases are worth to be pointed out~:
\begin{itemize}
 \item{ $\lambda =-1$}  \\
Using the variable $x = \cos(\phi)$, the vector space ${\cal V}$
can be re-expressed is the form
 \be
    {\cal V} = {\rm span} \{
    \cos(n \phi), \sin(n \phi), \cos((n-1)\phi),
    \sin((n-1)\phi), \dots
     \}
 \ee
 and the operators $S_a$ above can be expressed 
 in terms of trigonometric functions.
 Exemples of QES equations of this type were studied
 in \cite{uz} in relation with spin systems.
 \item{ $\lambda = k^2$ }
 Using the variable $x = {\rm sn}(z,k)$, (with   ${\rm sn}(z,k)$
 denoting the Jacobi elliptic function of modulus $k$, 
 $0 \leq k^2  \leq 1$), and
 considering the Lam\'e equation~:
 \be
 - \frac{d^2 \psi}{dz^2} + N(N+1)k^2 {\rm sn}^2(z,k)\psi = E \psi
 \ee
 It in known (see e.g. \cite{arscott})  that doubly periodic solutions
 exist if $N$ is a semi integer. 
 If $ N = (2n+1)/2$ these solutions are of the form
 \be
 \label{lame}
    \psi(z) = \sqrt{{\rm cn}(z,k) + {\rm dn}(z,k)}
    (p_n(x) + {\rm cn}(z,k){\rm dn}(z,k) p_{n-1}(x))
 \ee
where ${\rm cn}(z,k), {\rm dn(z,k)}$ denote the other Jacobi
elliptic functions.
The second factor of this expression is exactly an element
of the vector space under consideration.
The relations between the
doubly periodic solutions  of the Lam\'e equation
and QES operators was pointed out in \cite{bg}.
\end{itemize}
\subsection{Case $f(x) = \sqrt{(1-x)/(1-\lambda x)}$, $m=n$}
In this case again, the vector space ${\cal V}$ for $m=n$
is preserved by the three operators $S_a$ of Eq.(\ref{sop}) provided
$m=n$.
Two cases are worth considering, in complete paralelism 
with Sect. 2.3:
\begin{itemize}
\item{$\lambda = -1$.}
Using the new variable $x =\cos \phi$, and using the identity
$\tan(\phi/2) = \sqrt{(1-x)/(1+x)}$ the vector space
${\cal V}$ can be reexpressed is the form
\be
{\cal V} = {\rm span}\{
\cos(\frac{2n+1}{2} \phi),
\sin(\frac{2n+1}{2} \phi),
\cos(\frac{2n-1}{2} \phi),
\sin(\frac{2n-1}{2} \phi), \dots
\}
\ee
and exemples of QES operators having
solutions in this vector space are presented in \cite{uz}.
\item{$\lambda = k^2$.}
 Again, in this case, the variable  $x = {\rm sn}(z,k)$
 is usefull and the doubly periodic solutions 
 of the Lam\'e equation (\ref{lame})
 corresponding to $N=(2n+3)/2$ of the form \cite{bg}
 \be
    \psi(z) = \sqrt{{\rm cn}(z,k) + {\rm dn}(z,k)}
    ( {\rm cn}(z,k) p_n(x) + {\rm dn}(z,k) q_{n}(x))
 \ee
 provide  examples of QES solutions constructed in the space under
 consideration.
\end{itemize}

\section{Conclusions}
The problem considered in this paper 
enlarges the  framework
where QES operators are usually constructed
and offers a variety
of non trivial forms of the function $f(x)$.
Very recently, similar kinds of extensions of QES operators
were found \cite{guk},
although treated with a different orientation than the one
proposed here.

Several QES operators obtained in different contexts
\cite{uz,bg,dv} are recovered by our method in a unified way.
Other attempts
with different form of the function $f(x)$ turned out to
be trivial or to admit very involved (or very poor)
sets of preserving operators.
Other simple forms of the function $f(x)$ could definitely
be looked for.
The construction of QES operators preserving the space
${\cal V} = {\cal P}_n + f {\cal P}_m$
can be generalized in many directions. Namely   (i) to
functions of several variables, (ii) to the matrix case
(i.e to a space defined by
the direct sum two or more spaces of the type ${\cal V}$).
Apart from the algebraic problem of classifying the linear
differential operators preserving these vector spaces, 
the construction of new QES Schr\"{o}dinger operators
constitutes another interesting problem.
The quantum hamiltonians constructed in \cite{bh} are constructed
along the line  (ii) mentionned above.

\end{document}